\documentclass[showpacs,twocolumn,aps,pre,floatfix,superscriptaddress]{revtex4}
\usepackage{epsfig}
\usepackage{bm}

\begin{document}
\newcommand{\binc}[2]{B_{#1,#2}}

\title{Caging and mosaic lengthscales in plaquette spin models of
glasses}

\author{Robert L. Jack} \affiliation{Rudolf Peierls Centre for
Theoretical Physics, University of Oxford, 1 Keble Road, Oxford, OX1
3NP, UK}

\author{Juan P. Garrahan} \affiliation{School of Physics and
Astronomy, University of Nottingham, Nottingham, NG7 2RD, UK}

\begin{abstract}
We consider two systems of Ising spins with plaquette interactions.
They are simple models of glasses which have dual representations as
kinetically constrained systems.  These models allow an explicit
analysis using the mosaic, or entropic droplet, approach of the random
first-order transition theory of the glass transition.  We show that
the low temperature states of these systems resemble glassy mosaic
states, despite the fact that excitations are localized and that there
are no static singularities.  By means of finite size thermodynamics we
study a generalised caging effect whereby the system is frozen on
short lengthscales, but free at larger lengthscales.  We find that the
freezing lengthscales obtained from statics coincide with those
relevant to dynamic correlations, as expected in the mosaic view.  The
simple nucleation arguments of the mosaic approach, however, do not
give the correct relation between freezing lengths and relaxation
times, as they do not capture the transition states for relaxation.
We discuss how these results make a connection between the mosaic and
the dynamic facilitation views of glass formers.
\end{abstract}


\maketitle

\section{Introduction}

The aim of this paper is to study possible connections between two
perspectives on the glass transition \cite{Reviews}.  One is the
mosaic, or entropic droplet, view that follows from the random
first-order transition theory of Kirkpatrick, Thirumalai and Wolynes
(KTW) \cite{KTW,Xia-Wolynes}: a deeply supercooled liquid is viewed as
a patchwork of correlated mesoscopic regions which relax by
entropically driven nucleation events.  These ``mosaics'' are
characterised by a length scale that diverges on approach to a
temperature $T_K$, where a Kauzmann \cite{Kauzmann,Rome} entropy
crisis occurs.  The growth of static correlations, under certain
assumptions for droplet nucleation, leads to a divergence of
relaxation times of the Vogel-Fulcher kind \cite{Reviews} in three
dimensions.

The second, in principle very different, approach is based on the idea
of dynamic facilitation \cite{Fredrickson-Andersen,Palmer-et-al}:
glassiness is not due to any precipitous thermodynamics, but is a
consequence of effective constraints in the dynamics.  Here, the
central feature is dynamic heterogeneity \cite{DH,Other-theories}, and the
corresponding dynamic scaling, i.e., growing times are accompanied by
growing dynamic, but not necessarily static, correlation lengths
\cite{Garrahan-Chandler}.  In this view there are no finite
temperature singularities, and scaling properties are controlled by
zero temperature critical points where dynamic lengths and times
diverge \cite{Whitelam-et-al}.  The simplest models that realize this
perspective are kinetically constrained lattice models, such as the
facilitated spin models of Fredrickson and Andersen (FA) and Jackle's
East model \cite{Fredrickson-Andersen,East,Ritort-Sollich}.

Here we study lattice spin models with plaquette interactions
\cite{Lipowski,Newman-Moore,Garrahan-Newman,Garrahan} from the mosaic
or entropic droplet perspective.  The two models we consider are the
square plaquette model (SPM) \cite{Lipowski,Garrahan} and the
triangular plaquette model (TPM) \cite{Newman-Moore,Garrahan-Newman}.
These models have exact dual descriptions: one in terms of interacting
spin variables with standard single-spin Glauber dynamics; another in
terms of free excitations with dynamics subject to kinetic
constraints.  The dynamics of the SPM is like that of the
facilitated FA model \cite{Garrahan}, the dynamics of the TPM, like
that of the East model \cite{Garrahan-Newman}.

While plaquette models are realizations of the dynamic facilitation
scenario, they also allow for a detailed and explicit
analysis using the mosaic approach.  They therefore allow a direct
analysis of the similarities and differences between the facilitation and
the mosaic perspectives on glasses.  In this work, we apply to these
models the procedure recently suggested by Bouchaud and Biroli (BB)
\cite{Bouchaud-Biroli}, by which the mosaic lengthscale is estimated
from the partition function of finite droplets within a larger system.
We show that low temperature states do resemble mosaic states, and
that the associated lengths can be extracted using the BB procedure.
We also show that these thermodynamic lengthscales have the same
scaling as those extracted from bulk many-point static correlations,
and from multi-point dynamical correlation functions which measure
dynamic heterogeneity.  In the case of the SPM, the static analysis
gives two typical caging lengths, a feature not anticipated in BB.
Moreover, the relation between caging lengths and relaxation times in
these models is not that expected from the mosaic approach, as the
nucleation assumptions miss the relevant transition states and
overestimate the free-energy barriers to droplet melting.

The paper is organized as follows.  In section II we give a summary of
the arguments  of KTW and BB,  and make general  observations about the
applicability of this approach to the models studied in this work.  In
section III we apply the mosaic procedure to the SPM, and compare the
lengthscales obtained  with other relevant static  and dynamic lengths
of the  system.  In section  IV we repeat  the analysis for  the TPM.
Finally, in section V we give our conclusions.

\section{Background}
\label{sec:BB}

\subsection{Mosaics and their lengthscales}

The basic arguments underlying the BB procedure to compute a mosaic
lengthscale are the following \cite{KTW,Bouchaud-Biroli}.  

Imagine a glassy state in which an atom is prevented from moving by
the fact that its neighbours are fixed in position. This picture
is consistent only insofar as the neighbours of the original
atom are fixed by their neighbours, and so on. 
Consider now the original atom and its neighbours as a small droplet.  
One can compute
the probability that the droplet can relax, assuming that its
boundaries are fixed.  If this probability is small then
rearrangements require correlated motion over distances larger than
this droplet.  One must then increase the droplet size and recalculate
the probability of rearrangement.  
In a system with finite-ranged interactions, as the droplet is
made larger, we eventually cross over to a regime where the
rearrangement probability is large.  All larger droplets will be then
able to rearrange by sequential moves of this type.

This crossover defines the mosaic lengthscale
\cite{KTW,Bouchaud-Biroli}.  The result is a picture of a glassy state
in which droplets of small sizes are jammed, or frozen, by their
boundaries, whereas those with large sizes are unjammed, or melted. As
the temperature decreases, the typical lengthscale separating jammed
and unjammed droplets increases.  (The idea of using finite size
scaling to extract a lengthscale is not at all new, of course.  See
Refs.~\cite{Scheidler-et-al,Berthier} for applications to glass
formers.)

The above remarks can be made quantitative by way of three assumptions
\cite{KTW,Bouchaud-Biroli}:

(i) A droplet of size $\xi$ contains very many metastable states whose
 number scales as $e^{s_c(T) \xi^d}$, where $s_c$ is the
 configurational entropy density which vanishes at the Kauzmann
 temperature, $T_K$.

(ii) The probability of finding a droplet in a given state $i$, given
that its boundary conditions are in the state $\alpha$ is
\begin{equation}
p_{i\alpha} = \mathcal{Z}^{-1} e^{-\beta \left( \xi^d f_i +
\xi^{\theta}\Upsilon_{i\alpha} \right) } ,
\end{equation}
where $\mathcal{Z}$ is a partition sum, $f_i$ is the free energy
density of state $i$ and $\xi^{\theta}\Upsilon_{i\alpha}$ is the free
energy cost of matching the bulk state $i$ to the boundary state
$\alpha$.  The exponent $\theta$ should satisfy $0\leq\theta<d$ in a
short ranged system.  

(iii) Moves involving co-operative motion over a lengthscale $\xi$
occur on a timescale
\begin{equation} 
\tau \sim e^{\beta \Delta \xi^\psi} ,
\label{equ:tau_psi} 
\end{equation}
where $\Delta$ is a microscopic energy scale, and $\psi$ is some
exponent.  If we assume that these moves involve nucleation
events of one state in a background of an uncorrelated state then
the exponent $\psi$ is expected to obey $\psi\geq\theta$.

The lengthscale of the mosaic state can be extracted by comparing the
probability of a droplet state satisfying a frozen boundary condition,
with the total probability of all other droplet states with the same
frozen boundary \cite{Bouchaud-Biroli}.  That is, choosing a state $i$
such that $\Upsilon_{i\alpha}=0$, we define the crossover length
$\xi_*$ by
\begin{equation}
e^{-\beta f_i \xi_*^d} = \sum_{j\neq i} e^{-\beta \left( \xi_*^d f_j +
\xi_*^{\theta}\Upsilon_{j\alpha} \right) } .
\label{equ:mos_len}
\end{equation}
The right hand side is a sum over $e^{s_c \xi_*^d}$ terms, so we can
approximate it by
\begin{equation}
e^{-\beta f_i \xi_*^d} \simeq e^{- \left\{ \xi_*^d \left[ \beta f(T) -
   s_c(T) \right] + \xi_*^{\theta} \beta\Upsilon(T) \right\} } ,
\label{equ:mos_len_typ}
\end{equation}
where the sum over states $j$ was replaced with the typical number of
terms in the sum, $e^{\xi^d s_c(T)}$, multiplied by the typical weight
$e^{-\beta [\xi^d f(T)+ \xi^\theta\Upsilon(T)]}$.  We have included
explicitly the temperature dependence of $f$, $s_c$ and $\Upsilon$, as
they have been averaged over internal configurations of the droplet,
subject to the boundary condition $\alpha$.  As long as the free
energy of the well-matched state is typical, $f_i \simeq f(T)$, the
mosaic lengthscale is
\begin{equation}
\xi_* \sim \left( \frac{\beta \Upsilon}{s_c}
\right)^{\frac{1}{d-\theta}} .
\end{equation}
The configurational energy of the droplet is thought to vanish at some
finite $T_K$, which leads to the divergence of $\xi_*$ at that
temperature \cite{KTW}.

Combining these thermodynamical arguments, with assumption (iii) above
we get the typical relaxation time for a droplet of size $\xi_*$,
\begin{equation}
\tau \sim \exp(\beta \Delta \xi_*^\psi) .
\end{equation}
Droplets of length smaller than $\xi_*$ will not relax at all in this
timescale.  Droplets of length larger than $\xi_*$ relax by
combinations of moves over lengthscales of the order of $\xi_*$ in
times of the order of $\tau$. Hence, $\tau$ gives the typical rate of
rearrangements in the system \cite{xi-variable}.

\subsection{Plaquette models}
\label{sec:models}

The BB procedure outlined in the previous subsection is extremely
general.  However, analysing the partition function of any model in
sufficient detail to evaluate the length $\xi_*$, as defined in
Eq.~(\ref{equ:mos_len}), is rather difficult.  In the rest of this
work we apply the procedure to two models of Ising spins with
plaquette interactions in two dimensions.  

We will define the models in the next section, but we first make some
introductory remarks.  The first model is the square plaquette model
(SPM), in which spins are defined on a square lattice, and
interactions involve the quartets of spins that form the square
plaquettes of the lattice \cite{Lipowski,Garrahan}.  In the second
model, the spins occupy the sites of a triangular lattice, and
interactions are between triplets of spins on upward pointing
triangular plaquettes \cite{Newman-Moore,Garrahan-Newman}.  We refer
to it as the triangular plaquette model (TPM).  Both models have dual
representations \cite{Garrahan-Newman,Garrahan}: in a spin
representation they describe interacting spins with simple dynamical
rules (single spin-flips); in a ``defect'' representation they
describe thermodynamically independent excitations with more
complicated dynamical rules.  In this latter description they resemble
the kinetically constrained models advocated as effective models for
the glass transition in the dynamic facilitation approach
\cite{Garrahan-Chandler}.  In this sense, they interpolate between
models with microscopic degrees of freedom, and more phenomenological
models of ``mobility fields''.

The SPM and TPM have trivial thermodynamics: the excitations in the
models are localized and statically non-interacting, so the only
thermodynamic singularity is at zero temperature. In terms of our
assumption (i), above, this means that we must have $T_K=0$. It also
has implications on the assumptions of (ii) and (iii) about the
surface tension and barrier exponents $\theta$ and $\psi$.  Below we
show that the BB procedure can still be applied to extract caging
lengthscales from finite-size static properties.  However, while
relaxation times grow with growing mosaic lengths, the quantitative
dependence of assumption (iii) does not hold in the SPM and TPM, and
these relations need to be generalized.

\section{Square Plaquette Model}
\label{sec:pq}

\begin{figure} \begin{center}
\epsfig{file=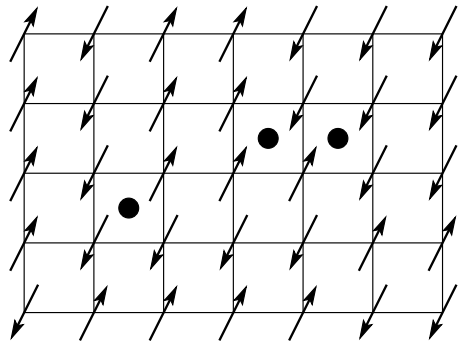,width=0.7\columnwidth}
\end{center}
\caption{Sketch showing spins (up or down) on vertices of a square
lattice, and defect variables (circled or blank) on the dual lattice
formed by the square plaquettes.  }
\label{fig:plaq_sketch}
\end{figure}

The SPM is defined by the Hamiltonian \cite{Lipowski,Garrahan}:
\begin{equation}
H = -(1/2)\sum_{xy} \left( \sigma_{xy} \sigma_{x,y+1} \sigma_{x+1,y}
\sigma_{x+1,y+1} -1 \right) ,
\end{equation}
where the $\{\sigma_{xy}\}$ are Ising spins, $\sigma=\pm 1$, and
$(x,y)$ indicates position on a square lattice.  The dynamical moves
are single spin-flips with Glauber rates.  If we write
\begin{equation}
p_{xy} = \sigma_{xy} \sigma_{x,y+1} \sigma_{x+1,y} \sigma_{x+1,y+1} ,
\end{equation}
then the partition function is simply a sum over the non-interacting
variables $p_{xy}$ (up to a non-extensive set of constraints on the
$p_{xy}$ from the boundary conditions on the spins). The variables
$p_{xy}$ are defined on the plaquettes of the square lattice, which
forms a dual square lattice, see Fig.\ \ref{fig:plaq_sketch}.
Plaquettes with $p_{xy}=-1$ cost an energy of unity, and are sparse at
low temperatures: we refer to them as defects.

The plaquette model has activated dynamics at low temperatures. That
is, there are very few (if any) possible transitions out of a typical
state into states with lower energy. Thus, low temperature states are
very close to inherent structures (on quenching to zero temperature, a
few local relaxations will take place, quickly leading to a state with
no available transitions). Thus it is an ideal model to investigate
the mosaic hypothesis: there are many metastable states, and these can
be easily identified. (In contrast, the statistics of the free energy
minima of an atomistic liquid or a spin glass are much harder to
probe).

The trivial bulk thermodynamics of the SPM are affected by freezing
the boundary spins of a finite droplet.  Consider the partition
function of a finite system of $L^2$ spins (assumed square for
convenience), with $(4L-4)$ frozen spins along the boundary.  The
state of the system is then defined by the configuration of the
$(L-2)^2$ remaining (bulk) spins, and its energy is given by (half of)
the sum of the $(L-1)^2$ defect variables, $p_{xy}$.  We see
immediately that there are more configurations of the plaquettes than
of the bulk spins: a partition sum over the bulk spins does not
contain all configurations of the plaquette variables.  Rather, there
are
\begin{equation}
n_\mathrm{plaqs} - n_\mathrm{spins} = (L-1)^2 - (L-2)^2 = 2L-3 ,
\end{equation}
constraints on the possible arrangements of the defects.

\begin{figure} \begin{center}
\epsfig{file=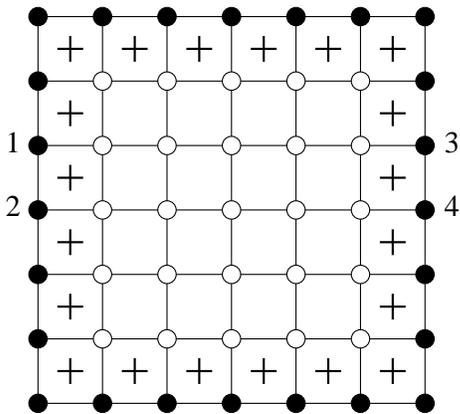,width=0.7\columnwidth} 
\end{center}
\caption{A droplet of size $L=7$. There are $(4L-4)=24$ frozen boundary
spins (black circles) and $(L-2)^2=25$ free bulk spins (white circles).
The energy is determined by the state of the $36$ defect variables that
sit on the dual lattice. This energy can be divided into a boundary
contribution that comes from the sites of the dual lattice marked
with $+$, and a bulk contribution from the remaining sites.
The partition sum is over the $2^{25}$ configurations of the bulk spins;
these correspond to $2^{25}$ of the $2^{36}$ configurations of the defect
variables. A given configuration of the defect variables contributes to
the partition sum if and only if it satisfies the $9$ 
independent constraints set by the 
frozen boundary spins. For example, one such constraint is that the
parity of the number of defects in the third row of the dual
lattice is the same
as the frozen spin combination $\sigma_1 \sigma_2 \sigma_3 \sigma_4$,
where the spins $\sigma_{1-4}$ are identified in the figure. }
\label{fig:pq_bulk_boundary}
\end{figure}

To understand the origin of these constraints, observe that flipping
any of the $(L-2)^2$ free spins preserves the parity of the number of
defects in any row or column of the square (dual) lattice. Therefore
the $2L-3$ constraints select the parity of the $L-1$ rows and $L-1$
columns of the dual lattice.  [Fixing the parities of all the rows
sets the parity of the total number of defects. Thus there are only
$L-2$ independent column parities, the final one being fixed by the
parity of the total number of defects.  The result is that there are
$2L-3$ independent constraints on the plaquettes, as required.]  These
arguments are illustrated in Fig.\ \ref{fig:pq_bulk_boundary}.

If we separate the energy of a configuration into a part coming from the
$(L-3)^2$ plaquettes in the bulk, and the $4L-8$ plaquettes along the
boundary then we may write the partition function for all $L^2$ spins
as
\begin{equation}
\mathcal{Z} = \sum_{i\alpha} z_{i\alpha}, \;\;\; \qquad z_{i\alpha} \equiv
e^{-\beta (n_i + n_{i\alpha})} ,
\label{equ:zia}
\end{equation}
where $i$ labels the configuration of the bulk spins and $\alpha$ that
of the boundary. The integer $n_i$ is the number of bulk plaquettes
that are excited, and $n_{i\alpha}$ is the number of boundary
plaquettes that are excited. Making contact with section~\ref{sec:BB},
we identify $n_i$ with the free energy $f_i$ and $n_{i\alpha}$ with
the surface energy $\Upsilon_{i\alpha}$.  Note that we have replaced
the sum in (\ref{equ:mos_len}) over inherent structures by a sum over
states, since the typical droplets under consideration cannot reduce
their energy by any single spin-flip.

The partition function in (\ref{equ:zia}) can be written in terms of
partial sums over boundary conditions $\alpha$, 
\begin{equation}
\mathcal{Z} = \sum_{\alpha} Z_\alpha, \;\;\; Z_\alpha \equiv \sum_i
z_{i\alpha} .
\end{equation}
Now imagine choosing a droplet in an infinite system.  In doing so we
find that its boundary state is $\alpha$.  The droplet lengthscale is
calculated by considering the contribution of that droplet to
$Z_\alpha$.   In other words, we need to estimate the expectation value
of $z_{i\alpha}/Z_\alpha$ as a function of droplet size.  This
expectation value is
\begin{equation}
\left\langle \frac{z_{i\alpha}}{Z_\alpha} \right\rangle_L =
 \frac{1}{\mathcal{Z}} \sum_{i\alpha} \frac{z_{i\alpha}^2}{Z_\alpha} =
 \sum_\alpha \frac{Z_\alpha}{\mathcal{Z}} \sum_i \left(
 \frac{z_{i\alpha}}{Z_\alpha} \right)^2 .
\label{equ:z_contrib}
\end{equation}
The second form for the expression above emphasises that it is an
average over the boundary conditions with their thermal weights,
$Z_\alpha$.

We define $\xi_*$ as the system size at which the above expectation
value ceases to be dominated by a single state,
\begin{equation}
\langle z_{i\alpha}/Z_\alpha \rangle_{\xi_*} \equiv (1/2) .
\label{equ:mos_len_Z}
\end{equation}

Since the droplet entropy is
\begin{equation}
S_\alpha = -k_B \sum_{i} (z_{i\alpha}/Z_\alpha) \log
(z_{i\alpha}/Z_\alpha) , 
\label{equ:droplet_entropy}
\end{equation}
Eq.\ (\ref{equ:mos_len_Z}) will be satisfied when $\langle S_\alpha
\rangle$ be of the order of $k_B$. In what follows we set $k_B=1$.
The entropy (\ref{equ:droplet_entropy}) measures the typical number of
states contributing to the partition sum with a fixed boundary.  It
must not be confused with the configurational entropy which measures
the number of states contributing to the overall partition sum.

\subsection{Finite size thermodynamics}

The partition function for a finite SPM with periodic boundaries was
calculated by Espriu and Prats \cite{Espriu-Prats}.  In
appendix~\ref{app:pq_series} we generalise their argument to allow for
a given state of $4(L-1)$ fixed boundary spins. The derivation is in
the spirit of a high temperature series, which can be resumed exactly
for this model since the bulk thermodynamics are those of a trivial
free lattice gas.  As mentioned above, the boundary conditions on the
spins constrain the parity of the number of defects in each row and
column of the dual lattice. If the number of rows with odd parity is
$r$ and the number of columns with odd parity is $r'$, then $Z_\alpha$
depends on the boundary conditions only through $r$ and $r'$.  Since
it is a symmetric function of these two integers, it is convenient to
define
\begin{equation}
m\equiv\mathrm{max}(r,r'), \qquad n\equiv|r-r'| , 
\end{equation}
so that $Z_\alpha = Z_{mn}$.  Note that $n$ is always even since $r$
and $r'$ (and therefore $m$) all have the parity of the total number
of defects in the droplet.  It is also convenient to parameterize
temperature by:
\begin{equation}
c \equiv e^{-\beta} ,
\end{equation}
which is related to the average concentration of excitations by
$\langle (1-p_{xy}) /2 \rangle = c/(1+c) \approx c$, the last
approximate equality being valid at low temperatures.

The low temperature behaviour of the plaquette model obeys scaling
relations \cite{Jack-et-al}.  Length and timescales diverge at low
temperatures as simple powers of $c$.  We therefore work in the
scaling regime, at leading order in $c$. In terms of droplet sizes,
this means that numerical values of $L$ may be large.  The relevant
indicators of droplet size will be $c L$ and $c L^2$. That is, we
assume $c \ll 1 \ll L$, but make no assumption on the absolute sizes
of $c L$ or $c L^2$.

We begin by considering fixed $L$ and very low temperatures: $cL^2 \ll
1$.  On picking a droplet from an infinite system, it is very likely
that it contains no defects.  The boundaries will then be such that
all rows and columns contain even numbers of defects.  There are
$2^{2L-1}$ such boundary conditions (one for each choice of spins,
say, in the bottom and leftmost sides of the box).  Tracing over the
bulk spins with each configuration of the boundary contributes
$Z_\alpha=Z_{00}\equiv Z_{m=0,n=0}$ to $\mathcal{Z}$:
\begin{equation}
\label{Z00cLL}
\mathcal{Z} = 2^{2L-1} [ Z_{00} + \mathcal{O}(cL^2) ] .
\end{equation}

\begin{figure}
\epsfig{file=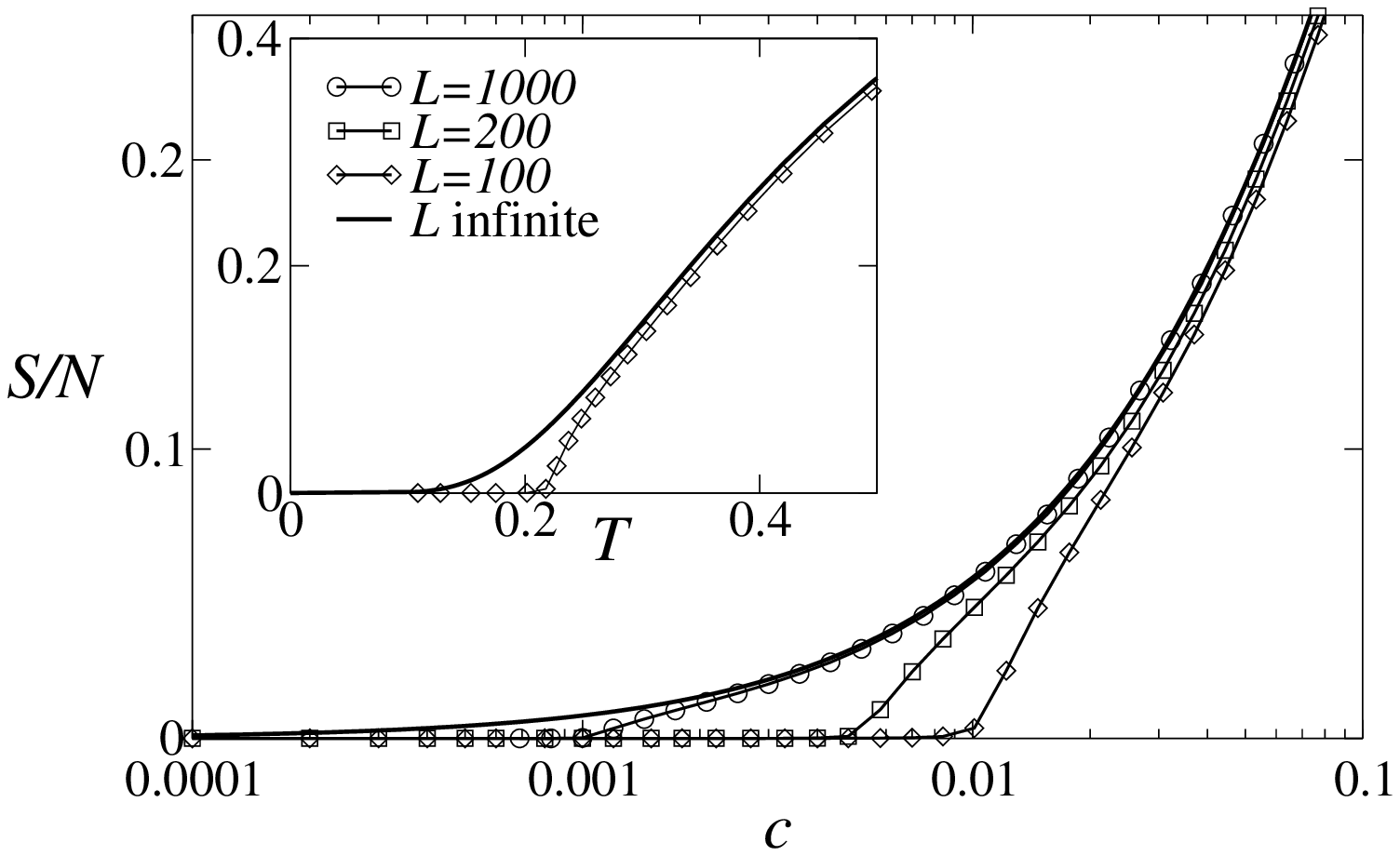,width=0.85\columnwidth}
\epsfig{file=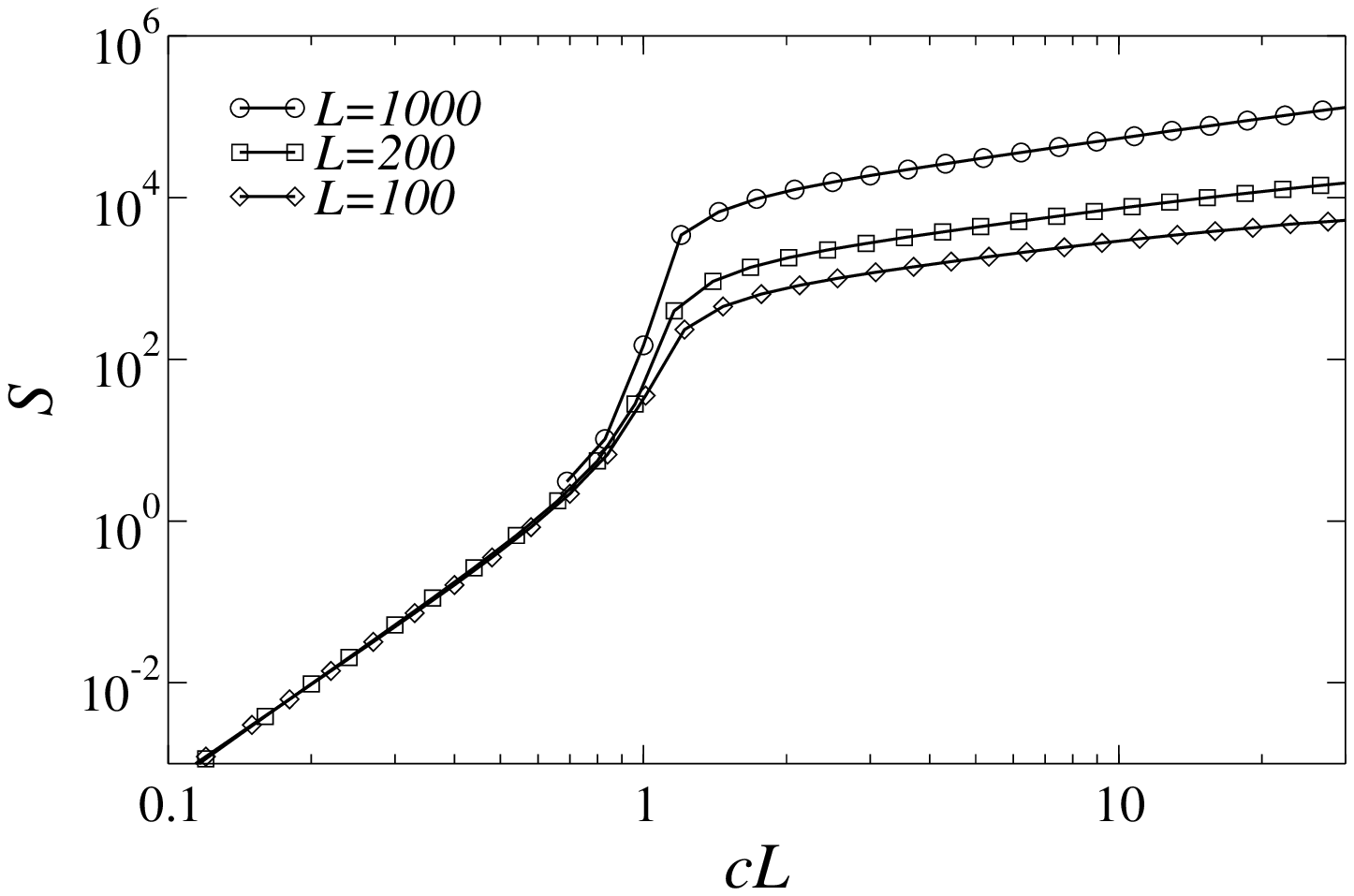,width=0.85\columnwidth}
\caption{(Top) Droplet entropy density $S_{00}/L^2$ versus $c \equiv
e^{-\beta}$ for droplets of different sizes $L$, with frozen boundary
conditions enforcing even numbers of defects in every row or column,
$m=n=0$.  The symbols are finite $L$ entropies calculated using the
expressions of appendix \ref{app:pq_series}.  The entropy is extensive
for $cL>1$.  On approaching $cL \sim 1$ from above it deviates from
the bulk entropy (solid line, $L\to \infty$), and becomes subextensive
for $cL<1$.  (Bottom) Plot of total entropy $S_{00}$ against $cL$
showing the (non-extensive) scaling predicted in (\ref{equ:Z00}). }
\label{fig:pbc_entropy}
\end{figure}

Now, if $m=n=0$ then the boundary of the droplet specifies its unique
ground state.  Further, simple counting arguments show that excited
states contain even numbers of defects $u\geq4$. Their Boltzmann
weights are $c^u$ and the degeneracy of the state with $u$ defects is
proportional to $L^{u}$. This means that:
\begin{equation}
\label{Z00cL} 
Z_{00}=1 + \mathcal{O}(cL) .
\label{equ:Z00} 
\end{equation}
This partition function is also relevant to periodic boundary
conditions, and is the one calculated in \cite{Espriu-Prats}.  In
Fig.\ \ref{fig:pbc_entropy} we show the droplet entropy $S_{00}=
\partial_T ( T \ln Z_{00})$, as a function of $c$ for different
droplet sizes $L$.  The top panel of Fig.\ \ref{fig:pbc_entropy} shows
that the droplet has an ``entropy crisis'' at $cL \sim 1$.  In the
bottom panel of Fig.\ \ref{fig:pbc_entropy} we see that the droplet
entropy scales with $cL$ when $cL\ll 1$, and is subextensive in this
regime.

From figure~\ref{fig:pbc_entropy}, it is clear that as the temperature
is increased, droplets with $m=n=0$ remain frozen until $cL\sim
1$. However, $Z_{00}$ dominates the droplet partition function only
for $cL^2 \ll 1$.  This means that \emph{typical} droplets are frozen
only for $cL^2\ll1$.  Between $c L \sim 1$ and $c L^2 \sim 1$ we must
consider $Z_{mn}$ with $m,n\neq0$.

\begin{figure}
\epsfig{file=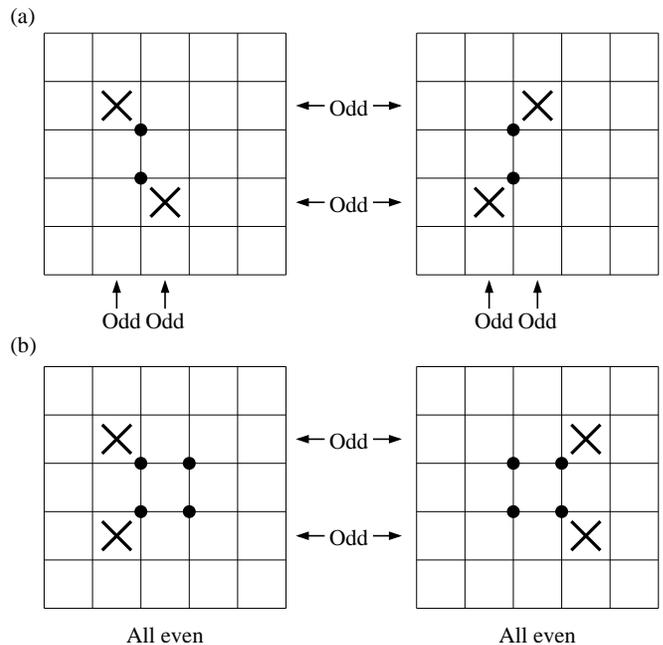,width=1.0\columnwidth}
\caption{Ground states of droplets ($L=6$) with different boundary
conditions.  Spins are on vertices on the square lattice and are not
shown. Excited plaquettes are marked with $\times$ symbols. (a)
Boundary conditions enforce two rows and two columns with odd numbers
of defects, so that $m=2$, $n=0$. There are two ground states that
differ by flipping the two spins identified with black circles. (b)
Boundary conditions enforce two rows with odd numbers of defects, but
all columns with even numbers: $m=2$, $n=2$. There are $(L-1)=5$
distinct ground ground states of which two are shown: they differ only
at the four marked spins}
\label{fig:bc_effect}
\end{figure}

\begin{figure}
\epsfig{file=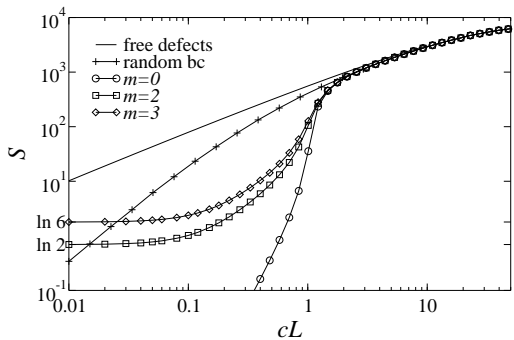,width=0.85\columnwidth}
\caption{Entropy as a function of $cL$ for $L=100$.  The full line is
  the bulk entropy of the SPM.  The open symbols are the entropies
  $S_{m0}$ for frozen boundaries with different $m$, as obtained from
  the finite size partition functions of appendix \ref{app:pq_series}.
  The + symbols give the entropy in the case where boundary conditions
  are generated randomly. }
\label{fig:Zm0_entropy}
\label{fig:entropy_c}
\end{figure}

The behaviour of droplets changes qualitatively in the case of ground
states containing two or more defects. As is clear from
Fig.\ \ref{fig:bc_effect}, if $m\geq2$ then there is more than one
state of the bulk spins that satisfies the boundary condition
perfectly.  Specifically, if $m=2$ then there are two ground states,
each of which contains two defects, and contributes $c^2$ to the
partition function. More generally, we find that
\begin{equation}
Z_{m0} \sim m!\, c^m [ 1 + \mathcal{O}(cL) ] .
\end{equation}
It follows that the zero temperature droplet entropy is now
$S_{m0}(T=0)=\log m!$, and so for these boundary states, we expect
typical values of
\begin{equation}
 z_{i\alpha} / Z_\alpha \sim (1/m!) + \mathcal{O}(cL) .
\end{equation}

While the droplets do have a feature in the entropy at $cL\sim1$, the
degenerate ground states all contribute equally to the partition sum
at low temperature.  Thus, as soon as droplets with $m\geq2$ dominate
$\mathcal{Z}$ then the they are thermodynamically melted according to
condition (\ref{equ:mos_len_Z}). Fig.~\ref{fig:entropy_c} shows that
this happens as soon as $cL^2 \sim 1$, as expected.  The melting
lengthscale is therefore,
\begin{equation}
\xi_* \sim c^{-1/2} .
\label{equ:pq_mos_len}
\end{equation}
 
It is clear that there are two crossovers in the droplet
thermodynamics as $c$ is increased. For $cL^2 \ll 1$ nearly all
droplets are frozen in true ground states of the system.  For $cL\gg
1$ the system behaves as the bulk.  In the intermediate regime $cL \ll
1 \ll cL^2$, each droplet is almost certainly in a state that
minimises the energy with respect to the boundaries; however, there
are many of these states for a fixed configuration of the boundary.
In this regime $m$ is about $cL^2$.  The low temperature droplet
entropy is then
\begin{equation}
S_{T=0} \simeq cL^2 \log cL^2 ,
\end{equation}
which is large compared to unity, but small compared to its bulk value
$cL^2\log 1/c$ (since $cL \ll 1$ implies $cL^2 \ll 1/c$).  We show
this explicitly in Fig.\ \ref{fig:entropy_c} where we average the
exact results for the entropy at given $m$ and $n$ over a distribution
of $mn$ that is sampled by Monte Carlo.

Thus far we have analyzed only states with $n=0$.  That is, the
boundary condition requires $m$ rows to have odd numbers of defects in
them, and it also requires $m$ columns to have the same property.  The
probability that $n=0$ is the probability that no two defects occupy
the same row or columns as each other,
\begin{equation}
P(n=0) = 1-\mathcal{O}(c^2 L^3) .
\end{equation}
This probability is close to $1$ as long as $c L^2 \ll \sqrt{L}$.

The melting length (\ref{equ:pq_mos_len}) can also be accessed through
static overlap functions, which will be important in our discussion of
dynamics.  Consider the self-overlap for a given boundary $\alpha$,
\begin{equation}
C_{0\alpha} = \frac{1}{(L-2)^{2}} \sum_{xy}
\left\langle  \sigma_{xy} \right\rangle_\alpha^2 ,
\label{equ:c0_replica}
\end{equation}
where we have explicitly summed over the spins of a droplet with given
boundary condition, since for a given realisation of the $\alpha$,
single spin correlations are finite and depend on position.  The
brackets $\langle \dots \rangle_\alpha$ denote a thermal average over
configurations of the bulk spins, with the boundary spins fixed.
$C_{0\alpha}$ should give the long time limit of the spin-spin
autocorrelation for a droplet with given boundary conditions.
Averaging over the boundary spins we get:
\begin{equation}
C_0 = \sum_\alpha Z_\alpha C_{0\alpha} ,
\label{equ:C0_replica}
\end{equation}
which is analogous to a disorder average.

\begin{figure}
\epsfig{file=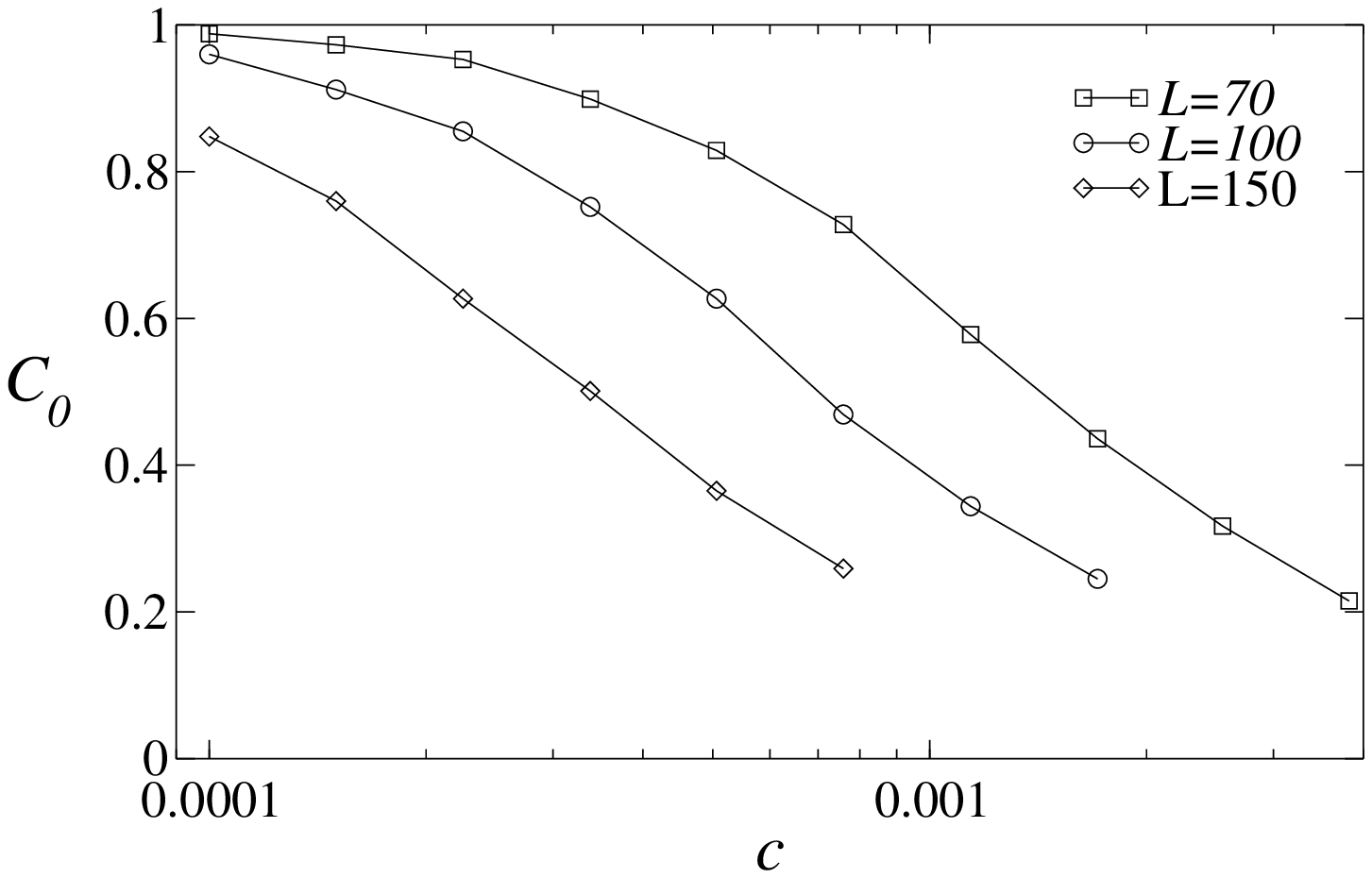,width=0.8\columnwidth}
\epsfig{file=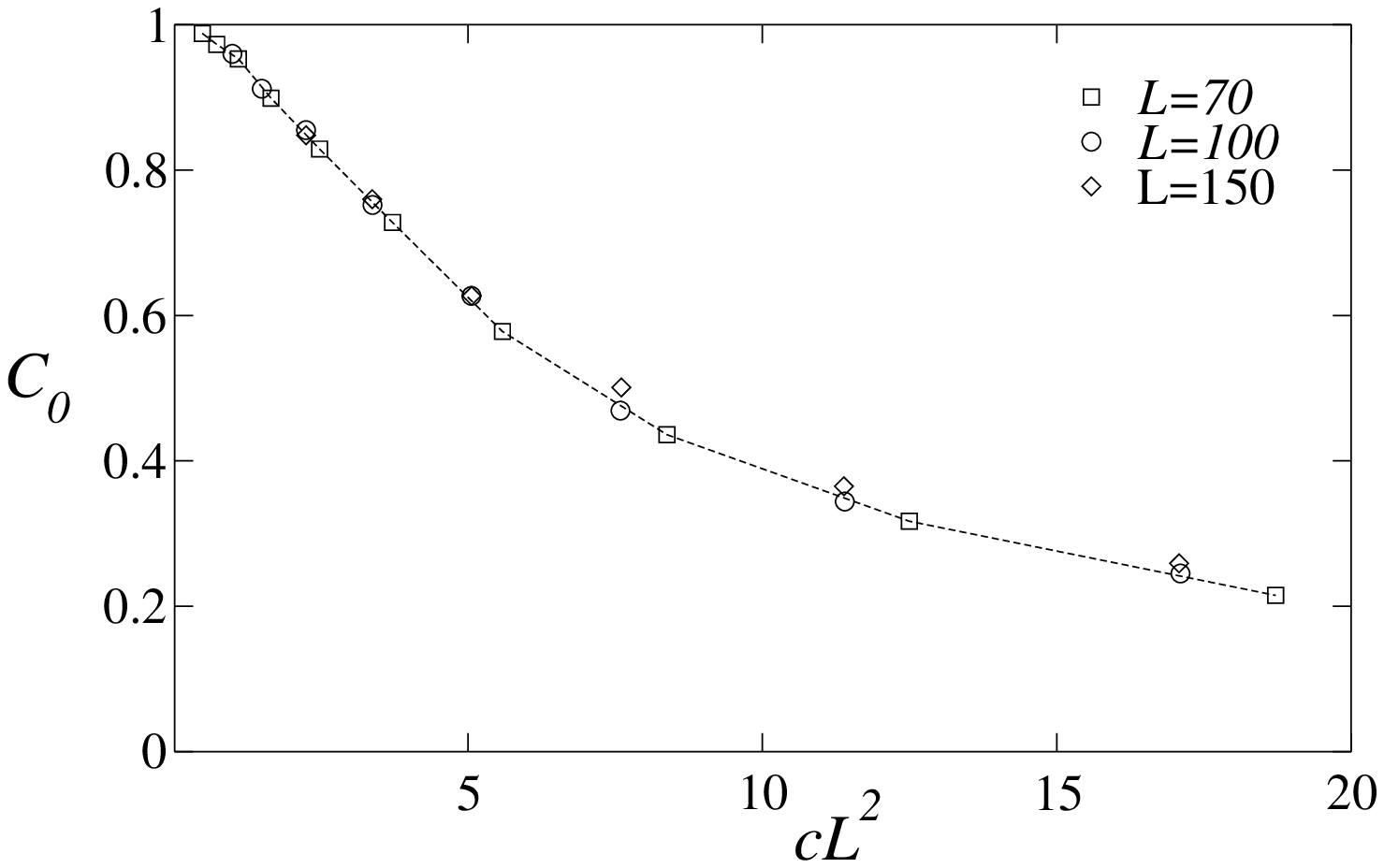,width=0.8\columnwidth}
\epsfig{file=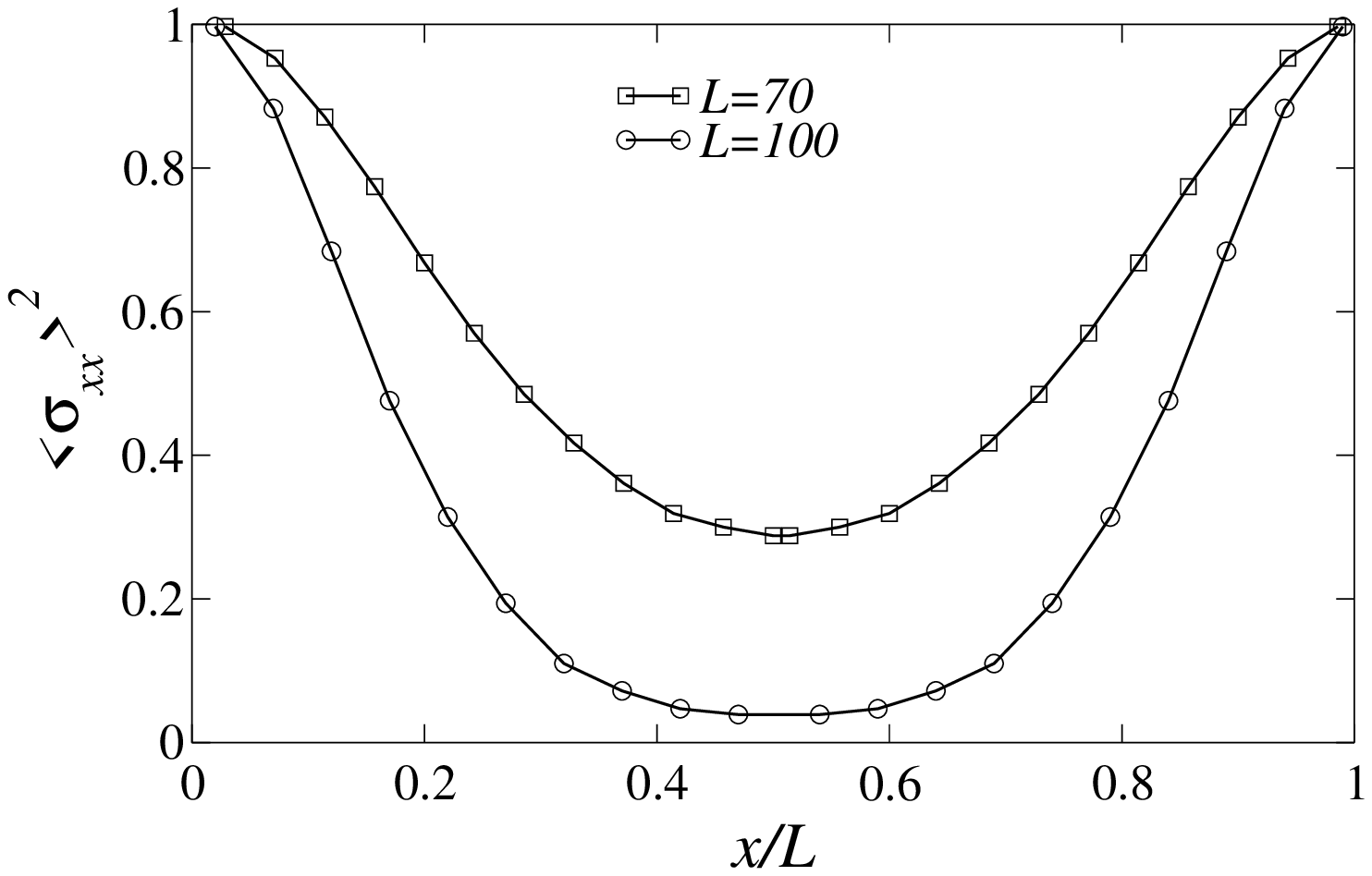,width=0.8\columnwidth}
\caption{Plots of the overlap function $\langle \sigma_{xy}
\rangle_\alpha^2$, averaged over boundary conditions $\alpha$ at
various temperatures. (Top) Average overlap $C_0$ as a function of
temperature. (Middle) Collapse of this data as a function of $cL^2$
(the dashed line is a guide to the eye). (Bottom) Spatial dependence
along the diagonal of the square droplet, showing strong correlations
near the boundary.  }
\label{fig:replica_correl}
\end{figure}

In Fig.\ \ref{fig:replica_correl} we show these overlaps.  The
correlator $\langle \sigma_{xy} \rangle_\alpha$ can be calculated in a
high temperature series: see appendix \ref{app:pq_series}.  The top
and middle panels of Fig.\ \ref{fig:replica_correl} show that $C_0$ is
a function of $c L^2$, which is close to $1$ for small $c L^2$, and
becomes small when $cL^2$ gets large.  This scaling also indicates
that $C_0$ will be small in the regime $cL \ll 1 \ll cL^2$.  The
bottom panel of Fig.\ \ref{fig:replica_correl} shows the spatial
dependence of $\langle \sigma_{xy} \rangle_\alpha$.  The spins near
the boundaries are strongly fixed by the boundary condition.  This
correlation decays away from the boundary with a lengthscale that
scales as $\xi_4 \sim c^{-1/2}$.  These observations for the overlaps
are consistent with thermodynamic melting of the droplet at $cL^2\sim
1$ found above.

\subsection{SPM and mosaic lengths}

We have established that the melting lengthscale $\xi^*$ scales as
$c^{-1/2}$ in the SPM, but that there is also a second important
thermodynamic crossover at $L\sim c^{-1} > \xi_*$.  We now compare the
results so far with the expectations of BB. The appearance of two
lengthscales was not anticipated in \cite{Bouchaud-Biroli}. We have
shown that the crossover at $cL^2\sim1$ comes from a small number of
states that are have similar weights in the droplet partition
function. We find that distinguishing these few specific states from
typical states with bulk energy $E$ is necessary when discussing the
droplet partition function $Z_\alpha$.

We use (\ref{equ:mos_len}) to define the mosaic lengthscale $\xi_*$.
The assumption of BB is that the lengthscale defined by
(\ref{equ:mos_len_typ}) will scale in a similar way.
Eq.~(\ref{equ:mos_len}) treats all states individually, whereas
(\ref{equ:mos_len_typ}) assumes that we can extrapolate the properties
of the partition sum from typical states at the relevant
temperature. We write the mosaic lengthscale extracted from
(\ref{equ:mos_len_typ}) as $\xi_{*,\mathrm{typ}}$: we find that
$\xi_*$ and $\xi_{*,\mathrm{typ}}$ do not coincide in the SPM.

More precisely, the average bulk energy of a droplet at a given
temperature is $c(L-2)^2$, and the configurational entropy at that
temperature is $c(L-2)^2 \ln(1/c)+(2L-1)\ln2$. If we replace the bulk
of the droplet by a new bulk state with the same energy, then the
typical boundary energy is $2(L-2)$ (half of the boundary plaquettes
will be excited, since there are no two-point spin correlations either
in the bulk or along the boundary). So from
Eq.~(\ref{equ:mos_len_typ}), we have $d=2$, $\theta=1$ and
$$\xi_{*,\mathrm{typ}}\sim c^{-1}$$ 

On the other hand, we showed in the previous section that the
Eq.~(\ref{equ:mos_len}) leads to
$$\xi_{*}\sim c^{-1/2}$$
The origin of this discrepancy is simply the degenerate ground
states of the droplet, whose boundary energy does not
take the typical value $\Upsilon(T)\sim L$, but rather
have $\Upsilon_{i\alpha}=0$. Thus they must be separated from the
sum before it is approximated:
\begin{eqnarray}
Z_\alpha &=& e^{-\beta f_i L^2} + \sum_{j\neq i, j\in \mathrm{gs}}
e^{-\beta f_j L^2} \nonumber \\ && + \sum_{j\neq i, j \notin
\mathrm{gs}} e^{-\beta f_j L^2 - \beta \Upsilon_{j\alpha}}
\label{equ:Z_decompose}
\end{eqnarray}
where the sum over $j\in\mathrm{gs}$ denotes a sum over states with
$\Upsilon_{j\alpha}=0$.  The final term in (\ref{equ:Z_decompose}) can
then be approximated to give a form resembling
Eq.~(\ref{equ:mos_len_typ}), but that term is irrelevant to droplet
melting in the SPM since the second term dominates the partition
sum for $cL \ll 1 \ll cL^2$.

We interpret this result as evidence that approximations such as those
leading to Eq.~(\ref{equ:mos_len_typ}) are rather dangerous. In
general, systems may possess significant numbers of states in the
partition sum with smaller than average boundary energy, and these may
be sufficient to destroy spin correlations in the infinite time limit
($C_0\to0$ when $L\ll \xi_{*,\mathrm{typ}}$), even if the droplet is
exploring only a small fraction of its metastable states.  Further, we
will show in the next section that if the droplet does melt in stages
then choosing whether to identify the bulk relaxation time as
$\tau(\xi_*)$ or $\tau(\xi_{*,\mathrm{typ}})$ requires investigation
of the specific system under study. In the SPM, we will find that
while complete destruction of of spin correlations at large times does
take place for $L\ll \xi_{*,\mathrm{typ}}$, the timescale for this
relaxation is much longer than the bulk relaxation time. However, this
effect cannot be inferred from thermodynamic arguments: it is a
kinetic effect.

To end this discussion, we note that while the condition $cL^2\sim 1$
leads to rather small droplets, the configurational entropy of a
droplet of this size is $cL^2 \log(1/c)\gg1$. BB comment that the
vanishing of the configurational entropy leads to a trivial lower
bound on the mosaic lengthscale which in this case is: $\xi_*^2 >
[c\log(1/c)]^{-1}$. Thus we see that despite the condition that frozen
droplets contain at most one defect, the freezing lengthscale is still
much larger than its trivial lower bound.

\subsection{Relaxational dynamics}

In this subsection we consider the relaxation of the droplets studied
thermodynamically above.  The relevant correlation function is
\begin{equation}
C_\alpha(t) = 
\frac{1}{(L-2)^2}
\sum_{xy\in \hbox{bulk}} \langle \sigma_{xy}(0) \sigma_{xy}(t)
\rangle_\alpha
\end{equation}
where the average is over realisations of the Markov chain that
describes the time evolution of steady state of the system (with fixed
boundary state $\alpha$). We also define the average of $C_\alpha(t)$
over boundary conditions
\begin{equation}
C(t) = \sum_\alpha Z_\alpha C_\alpha(t)
\end{equation}
The long time limit of $c_\alpha(t)$ is given by the overlap
$c_{0\alpha}$.

\begin{figure}
\epsfig{file=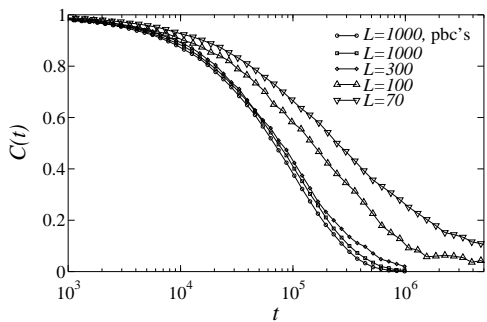,width=0.8\columnwidth}
\caption{Simulations of the SPM at $c=0.01$ for various system sizes
and frozen boundary conditions.  The relaxation time increases
significantly as $cL$ decreases through unity, but the long time limit
of $C(t)$ would still appear to vanish since $cL^2 \gg 1$.  We also
show relaxation for $L=1000$ and periodic boundary conditions.  The
relaxation can be well-fitted by an exponential for large $L$.  At
smaller $L$ the decay is slower than exponential.  }
\label{fig:pq_Ct_cL}
\end{figure}

\begin{figure}
\epsfig{file=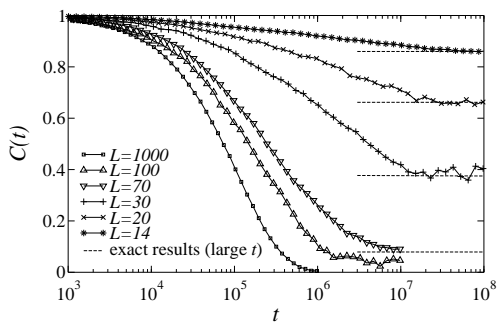,width=0.8\columnwidth}
\caption{Simulations of the SPM at $c=0.01$ for various
system sizes and frozen boundary conditions.  The long time limit of
$C(t)$ increases from zero as $cL^2$ becomes of order unity.}
\label{fig:pq_Ct_cL2}
\end{figure}

>From Monte Carlo simulations, it is clear that the relaxation of the
system slows down as the system size is decreased towards $1/c$: see
Fig.\ \ref{fig:pq_Ct_cL}.  From our discussions of the static single
point correlator [Eq. (\ref{equ:c0_replica}) and
Fig.\ \ref{fig:replica_correl}] we expect the long time limit of this
function to approach zero even for $cL \ll 1 \ll cL^2$, although this
is hard to show in simulations since simulations with large numbers of
defects at small temperatures are very computationally expensive.

Moving towards the regime with $cL^2 \sim 1$, we see that the
relaxation of the system is very slow, and that there are large
correlations that persist to the infinite time limit: see Fig.\
~\ref{fig:pq_Ct_cL2}. We can make a link between static and dynamic
quantities by evaluation of the long time limit of these functions
using the high temperature series (cf.\ Fig.\
~\ref{fig:replica_correl}).

Thus we have shown how the two lengthscales associated with the glassy
mosaic state affect the dynamics of the system.  As the temperature is
reduced, the relaxation slows for $cL\sim1$, before stopping
completely at $cL^2\sim1$. The separation of these lengthscales was
not anticipated in \cite{Bouchaud-Biroli}: we have more to say on this below.

Before ending our discussion of dynamics, we make two further
points. Firstly, we note that it is clear from Ref.~\cite{Jack-et-al}
that the bulk relaxation is caused by correlated motion of defects
over distances $\xi_\mathrm{dyn} \sim c^{-1}$. This coincides with the
system size above which the system relaxes as the bulk.  Secondly, it
is also known \cite{Espriu-Prats,Jack-et-al} that correlated motion
for defects over a distance $l$ takes place on a timescale $\tau \sim
l e^{2\beta}$ so we see that the analogue for equation
(\ref{equ:tau_psi}) is
\begin{equation}
\tau \sim e^{2\beta+\log \xi}
\label{equ:pq_tau_xi}
\end{equation}
The exponent depends logarithmically on $\xi$ with a prefactor that
does not depend on temperature: apart from the constant activation
energy, the free energy barriers to motion are purely entropic in
origin. The entropic nature of the barriers explains why relaxation of
the SPM model diverges only in an Arrhenius fashion ($\tau\sim
e^{3\beta}$) despite the co-operativity of its dynamics.

\subsection{SPM summary}

It is clear from the results of this section that, even in a simple
model like the SPM, the situation regarding caging and mosaic
lengthscales is already more complicated than what the appealingly
simple and general arguments of \cite{Bouchaud-Biroli} would predict.  As we
increase the size of droplets with frozen boundaries, there are two
crossovers.  Static spin-spin correlations vanish at at $c \sim
L^{-2}$, but neither the droplet entropy, nor its dynamical
correlation functions reach their bulk forms until the much higher
temperature $c\sim L^{-1}$.  We note that these two lengthscales were
also identified in the discussion of static and dynamic lengthscales
in Ref.~\cite{Jack-et-al}.  To be precise, the static four-spin
correlator $\langle \sigma_{00} \sigma_{0r} \sigma_{r0} \sigma_{rr}
\rangle$ decays on a lengthscale $\xi_4^\mathrm{stat} \sim c^{-1/2}$,
but the dynamic four point correlator decays on a length that scales
as $\xi_{2,2} \sim c^{-1}$.  There is also a static lengthscale that
mirrors the scaling of $\xi_{2,2}$ which is derived from fluctuations
of two point correlations \cite{Jack-et-al}.

Despite these complications, a general picture like that of
\cite{Bouchaud-Biroli} does seem to be applicable to the SPM, in the
sense that droplet melting does occur, even if it happens in
stages. As the temperature decreases towards the Kauzmann singularity
at $T=0$, the mosaic lengthscale increases, and the relaxation time
increases according to (\ref{equ:pq_tau_xi}).  This increase follows a
simple Arrhenius law because only the entropic barriers depend on
$\xi_*$.  This suggests that one should replace (\ref{equ:tau_psi}) by
the more general parametrization
\begin{equation}
\tau \sim f_s(\xi) e^{\beta f_u(\xi)}
\label{equ:tau_barriers}
\end{equation}
where $f_s(\xi)$ describes the entropic part of the barrier and
$f_u(\xi)$ the energetic part. At small temperatures then we expect
the energetic part to dominate, except in the case where $f_u(\xi)$ is
independent of $\xi$, as is the case for the SPM.

It is at this point at which our discussion of dynamics diverges from
that of KTW and BB: since the (non-perturbative) dynamical moves are
not nucleation events in this model, then the free energy barrier is
not set by a transition state with a critical droplet of the new
phase.  Rather, the relaxation proceeds by less expensive
rearrangements that occur over a lengthscale $\xi_{*,\mathrm{typ}}$,
as was discussed in detail in Ref.\cite{Jack-et-al}.

\section{Triangular plaquette model}
\label{sec:triag}

In this section we compare the results we found above for the SPM,
with a similar model of Ising spins: the triangular plaquette model,
or TPM \cite{Newman-Moore,Garrahan-Newman}. This model has a
super-Arrhenius divergence of the relaxation time at low temperatures,
and is therefore a fragile model, as compared to the SPM, which is
strong.  Moreover, dynamical correlations do not have the strong
anisotropy of the SPM \cite{Jack-et-al}.  The calculation of the
partition function, however, is considerably more involved, so our
results below are less detailed than those for the SPM.

The TPM is defined by the Hamiltonian
\cite{Newman-Moore,Garrahan-Newman}:
\begin{equation}
H = -(1/2) \sum_{xy} \sigma_{xy} \sigma_{x+1,y} \sigma_{x,y+1} .
\end{equation}
where as before $\{\sigma_{xy}\}$ are Ising spins on a square lattice
and the dynamics are is spin-flips with Glauber rates.  If we define
the plaquette variables
\begin{equation}
q_{xy} = \sigma_{xy} \sigma_{x+1,y} \sigma_{x,y+1}
\end{equation}
we again find that the partition sum over the spins reduces to a
product over $q_{xy}$, that are independent up to a non-extensive set of
constraints from the boundary conditions \cite{Newman-Moore,Garrahan}.

\begin{figure}
\epsfig{file=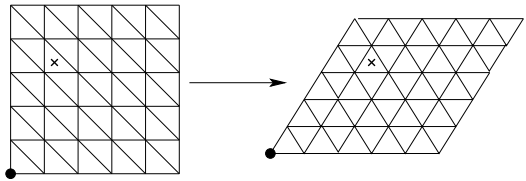,width=0.8\columnwidth}
\caption{Sketch showing relation between square and rhomboid droplets
($L=6$). The spins are on the intersections of grids; $\sigma_{11}$ is
marked by a filled circle. The $\times$ symbol marks the plaquette
interaction $q_{24}=\sigma_{24}\sigma_{25}\sigma_{34}$. Rotations of
$120^\circ$ leave both Hamiltonian and triangular lattices
invariant. However, note that the boundaries of the droplets are not
all equivalent, since the rhombus shape is not invariant under this
rotation.}
\label{fig:rhomb}
\end{figure}

We note that defining this model in this way leads to a Hamiltonian
that does not possess the square symmetry group of the lattice: it is
more intuitive to deform the square lattice into a triangular one. The
interactions are then around upward pointing triangular plaquettes
(see Fig.\ \ref{fig:rhomb}). The symmetries of the Hamiltonian are
then translations by lattice vectors, and $120^\circ$ rotations
($60^\circ$ rotations leave the lattice invariant but flip downward
pointing triangles to upward pointing ones, so they change the
Hamiltonian).

The analogy of the square droplet of Fig.\ \ref{fig:plaq_sketch} is a
rhombus shape in the triangular representation. The arguments for
counting bulk and boundary states are the same as those of the
plaquette model.  However, identifying the $2^{2L-1}$ degeneracies of
each state and the nature of the $(2L-3)$ constraints is considerably
more involved.

\subsection{Static spin-spin correlations and constraints}

The key to understanding static properties of the triangular plaquette
model is to realise that
\begin{equation}
\langle q_{i_1,j_1} q_{i_2,j_2} \dots q_{i_n,j_n} \rangle =
	[\tanh(\beta/2)]^n . 
\end{equation}
All multi-spin correlators that can be written in this form are
finite. Further, all other multi-spin correlators are zero
\cite{Garrahan}. While one and two point static spin correlations
vanish,
\begin{equation}
\langle \sigma_{xy} \rangle = 0, \;\;\; \langle \sigma_{xy}
\sigma_{x'y'} \rangle=\delta_{x,x'} \delta_{y,y'} ,
\end{equation}
specific three point correlators are non-zero,
\begin{equation}
\langle \sigma_{xy} \sigma_{x+1,y} \sigma_{x,y+1} \rangle =\langle
q_{xy} \rangle = \tanh(\beta/2) .
\end{equation}

\begin{figure}
\epsfig{file=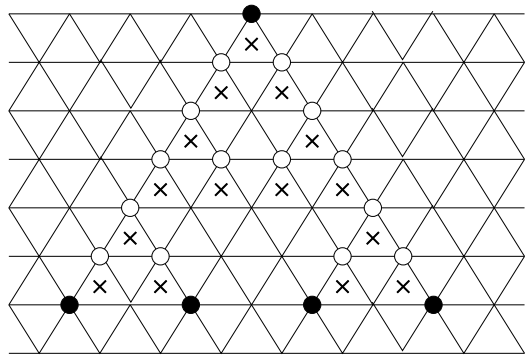,width=0.8\columnwidth}
\caption{Illustration of Eq.\ (\ref{equ:ST_correl}) with $k=6$. The
product of the five spins marked by filled circles is given by the
product of the fifteen plaquette variables marked with crosses. Both
the marked defects and the spins marked with circles (filled
\emph{and} empty) form parts of Sierpinski triangles: they are marked
if and only if the relevant entry of Pascal's triangle is odd.}
\label{fig:ST}
\end{figure}

A useful general relation is that
\begin{equation}
\sigma_{xy} \prod_{r=0}^k \sigma_{x+r,y-k}^{B_{k,r}} =
\prod_{k'=1}^k \prod_{r=0}^{k'-1} q_{x+r,y-k'}^{B_{k',r'}} ,
\label{equ:ST_correl}
\end{equation}
where the $B_{nr}\equiv n!/[r!(n-r)!]$ are binomial coefficients.
Since the $q$'s and the $\sigma$'s are Ising variables, only the
parities of the binomial coefficients are relevant.  In terms of
correlation functions, this relation implies that there is a finite
multi-spin correlator connecting $\sigma_{xy}$ and various spins in
row $x-k$ of the lattice.

This correlator is related to the first $k$ rows of Pascal's triangle
(PT). The correlator connects the spins corresponding to entries of
the $k$th row of the PT that have odd parity. Its value is given by
$\tanh(\beta/2)^p$ where $p$ is the number of entries in the first
$k-1$ rows of the PT that have odd parity. See Fig.\ \ref{fig:ST} and
the associated caption. Now, the typical number of odd entries in a
Pascal's triangle of linear size $l$ is $l^{d_\mathrm{f}}$ with
$d_\mathrm{f}=\log_2(3)$.  It follows that typical correlations over a
lengthscale $r$ scale as
\begin{equation}
\tanh(\beta/2)^{(r^{d_\mathrm{f}})} \sim 
\exp\left[-2(rc^{1/d_\mathrm{f}})^{d_\mathrm{f}}\right] ,
\end{equation}
from which we infer that static correlations decay on a lengthscale
\begin{equation}
\xi_\mathrm{stat}\sim c^{-1/d_\mathrm{f}} .
\end{equation}
In the case of finite droplets there must be $(2L-1)$ symmetry
operations that leave the energy of a droplet invariant, but require
the boundary spins to change (recall Fig.\
\ref{fig:pq_bulk_boundary}).  Furthermore, fixing the boundary spins
imposes $(2L-3)$ constraints on the allowed configurations of the
plaquette variables.  The form of these symmetries and constraints are
given explicitly in appendix~\ref{app:triag_constraints}.

The SPM was simple to analyse since the constraints are in a form
where each plaquette appears in exactly two constraints.  In that case
we can make a geometrical interpretation of those constraints.  In the
TPM the consequences of imposing the constraints is less clear.  We
therefore proceed directly to a discussion of dynamic properties, in
an attempt to access the static properties indirectly.

\subsection{Dynamic spin correlations}

Relaxation in the TPM is frustrated by a hierarchy of energy
barriers\cite{Newman-Moore,Garrahan-Newman}: co-operative motion over
a distance $\xi$ requires the crossing of an energy barrier of a
height $\log_2 \xi$.  This behaviour is reminiscent of the East model
\cite{East}.

\begin{figure}
\epsfig{file=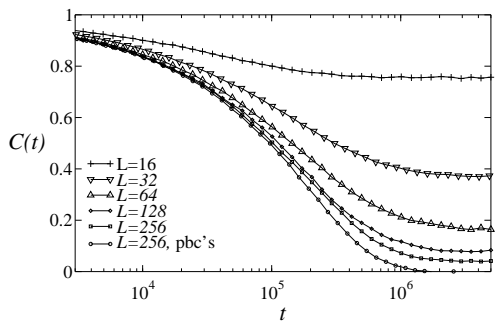,width=0.8\columnwidth}
\caption{Relaxation of finite systems at inverse temperature $\beta=4$
($c=0.018$).  Notice that the long-time limit of $C(t)$ is changing,
but that the relaxation time is approximately constant.  Compare to
Fig.\ \ref{fig:pq_Ct_cL2}, where the relaxation time increases
sharply as $C(t \to \infty)$ increases.}
\label{fig:triag_ct_small}
\end{figure}

In order to understand the scaling of the mosaic length, we
investigate the scaling of $C(t)$ in finite droplets: see
figure~\ref{fig:triag_ct_small}. A lower bound on
the freezing lengthscale is set by the elementary condition $cL^2 \sim
1$, since all states with single defects have different boundary
conditions. This would imply that the long time limit of the
autocorrelation, $C_0$ should be a scaling function of $cL^2$, as in
the SPM, Fig.\ \ref{fig:replica_correl}.

\begin{figure}
\epsfig{file=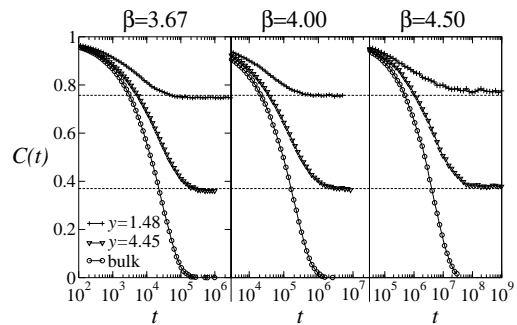,width=0.8\columnwidth}
\caption{Relaxation at different temperatures showing bulk results
(circles) and two different values of $cL^{d_\mathrm{f}}$.  The actual
system sizes used are (left) $L=13,26$; (middle) $L=16,32$; (right)
$L=26,44$.  The dashed lines are guides to the eye showing that the
long time limit scales as a function of $cL^{d_\mathrm{f}}$ only
(there are weak deviations from scaling which we attribute to
subleading corrections in $c$ and $1/L$).}

\label{fig:triag_ct_scaling}
\end{figure}

However, from the discussions of static correlations above, it is
clear that the physics of the triangular plaquette model is
intrinsically related to the parities of binomial co-efficients, and
hence to Sierpinski's triangle. As mentioned above, this structure has
a fractal dimension of $d_\mathrm{f}=\log_2(3)\simeq1.585$. In Fig.\
\ref{fig:triag_ct_scaling} we show relaxation data at various
temperatures with constant values of $cL^{d_\mathrm{f}}$. We see that
$C_0$ does seem to depend only on this combination. The conclusion is
that the mosaic lengthscale scales as
\begin{equation}
\xi_* \sim c^{-1/d_\mathrm{f}} .
\end{equation}
This relation indicates that typical droplets melt when they contain
of the order of $c^{1-2/d_\mathrm{f}}\sim c^{-0.26}$ defects; below
this number there no co-operative mechanisms allowing the droplet to
rearrange.  This should be compared with the SPM, in which droplets
with $L\sim\xi_*$ contain of the order of one defect, but those at the
dynamical crossover ($cL\sim 1$) contain of the order of $c^{-1}$
defects.

\begin{figure}
\epsfig{file=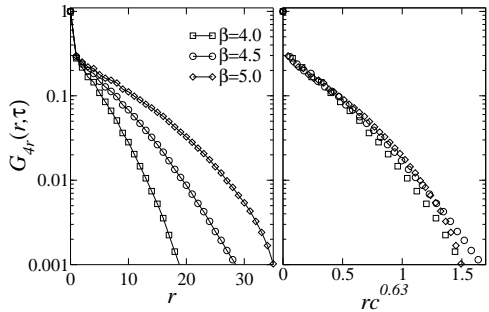,width=0.8\columnwidth}
\caption{Circular average of the normalised four point dynamic
correlator. The system size is $L=128$ with periodic boundaries, which
is large enough to avoid finite size effects.  (Left panel) Three
different temperatures, with times chosen so that $C(t)=0.5$.  (Right
panel) Collapse of the data of the left panel on rescaling of length
as $rc^{1/d_\mathrm{f}}$.}
\label{fig:triag_g4_scaling}
\end{figure}

An alternative measure of the co-operativity of the dynamics is given
by the bulk ``four-point'' correlation function (see
e.g. \cite{Toninelli-et-al})
\begin{equation}
\tilde G_4(xy;t) = \langle \sigma_{xy}(t) \sigma_{00}(t) \sigma_{xy}(0)
\sigma_{00}(t) \rangle .
\end{equation}
This correlation function measures spatial correlations in the single
site autocorrelation function.  It is convenient here to work with the
connected part of this function, normalised according to
\begin{equation}
G_4(xy;t) = \frac{\tilde G_4(xy;t) - 
\langle \sigma_{00}(t) \sigma_{00}(0) \rangle^2}
{1-\langle \sigma_{00}(t) \sigma_{00}(0) \rangle^2} .
\end{equation}
In the triangular plaquette model, the function $G_4$ has a
complicated spatio-temporal structure, again related to Sierpinski's
triangle: we postpone a detailed discussion of this structure to a
later work. However, the circular average of this function, which we
denote by $G_{4r}(r,t)$ decays smoothly with distance (at a fixed
time). This is shown in Fig.\ \ref{fig:triag_g4_scaling}, from which
it is clear that the lengthscale over which dynamical correlations
occur also scales as
\begin{equation}
\xi_\mathrm{dyn} \sim c^{-1/d_\mathrm{f}} .
\end{equation}

\begin{figure}
\epsfig{file=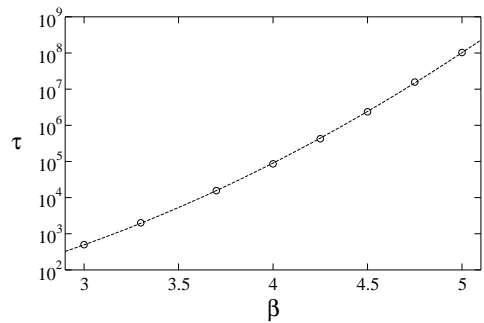,width=0.8\columnwidth}
\caption{Data showing relaxation time for large systems as a function
of inverse temperature, $\beta$. The relaxation time is defined by
$C(\tau)=0.5$. The error bars are smaller than the symbols shown. The
dashed line is a fit to the form $\ln \tau = a_0 + a_1\beta +
q\beta^2/\ln 3$.  Good fits are also possible with different
coefficients for the quadratic term.}
\label{fig:triag_tau}
\end{figure}

We have used the dynamics of finite droplets to identify a single
mosaic lengthscale in the triangular plaquette model, that scales as
$c^{-1/d_\mathrm{f}}$. Further, we know \cite{Garrahan-Newman} that
energy barriers to motion over a lengthscale $l$ require an activation
energy of the order of $\log_2 l$. At very low temperatures, we
therefore expect the bulk relaxation time at very low temperatures to
scale as 
\begin{equation}
\tau \sim \exp \left( \beta\log_2 \xi \right) \sim \exp \left( \beta^2
/ \ln 3 \right) .
\label{equ:triag_logtau}
\end{equation}
Note that Ref.\ \cite{Garrahan} predicted $\exp(\beta^2/\ln4)$ based
on the assumption that $\xi\sim c^{1/2}$.  Fig.\ \ref{fig:triag_tau}
shows a fit of the relaxation time of $C(t)$ with the function $\ln
\tau = a_0 + a_1\beta + \beta^2/\ln 3$.  The constant and linear terms
come from entropic and dynamical effects which become irrelevant at
very low temperatures where the energy barriers on $\xi$ dominate, see
Eq.~(\ref{equ:tau_barriers}).  The fit in the figure is consistent
with (\ref{equ:triag_logtau}), although the range of $\beta$ that we
consider is too small to confirm it beyond doubt.

\subsection{TPM summary}

The thermodynamics of finite droplets in the triangular plaquette
model are not sufficiently simple to allow an analysis of the droplet
entropy and hence determination of the mosaic lengthscale according to
the criterion (\ref{equ:mos_len}). However, it appears that the long
time limit of the single spin autocorrelation function approaches
unity as $cL^{\log_2 3}$ gets small.  If $C_0$ approaches unity then
it seems reasonable to assume that droplets are effectively trapped in
a single state that dominates the partition sum, and that this should
be an equivalent criterion to (\ref{equ:mos_len}).

We find that the mosaic lengthscale extracted in this way ($\xi_*$),
scales in the same way as the static lengthscale for multi-spin
correlations ($\xi_\mathrm{stat}$), and as the dynamical lengthscale
extracted from the four point correlator ($\xi_\mathrm{dyn}$). This
scaling is evidence for the presence of a single lengthscale in this
model. This is in contrast to the TPM, and more in line with
\cite{Bouchaud-Biroli}.

On the relation of dynamics to statics, the analogue of
(\ref{equ:tau_psi}) seems to be
\begin{equation}
\tau \sim f_s(\xi) e^{\beta \log_2 \xi}
\end{equation}
The increase of the energy barriers with $\xi$ is again slower than
the power law predicted by \cite{Bouchaud-Biroli}, but the assumption of an
increasing function is robust. The absence of a power law dependence
again indicates that the system relaxes by a more efficient mechanism
than nucleation of a new droplet in the background of an uncorrelated
state.

\section{Conclusions}
\label{sec:conc}

We have successfully applied the procedure of
Ref.~\cite{Bouchaud-Biroli} to two simple models of interacting Ising
spins.  We find that the concept of ``caging'' whereby the states of
individual spins are strongly constrained by frozen spins surrounding
them applies well in these models.  We were able to study the
breakdown of this caging effect by considering droplets of increasing
size.

In the square plaquette model (SPM), the caging breaks down in two
stages as the droplet size increases. For $L\ll c^{-1/2}$ the droplet
is frozen. For $c^{-1/2} \ll L \ll c^{-1}$ the droplet will eventually
relax, but this relaxation is slower than that of the bulk
system. Finally, for $L\gg c^{-1}$ the relaxation time coincides with
that of the bulk.  The correlation lengths controlling these two
crossovers have the same scalings as the two separate lengths
identified in this model from four point static and dynamic
correlations\cite{Jack-et-al}. The two stage droplet melting in this
model is more complicated than the simple scenario of BB.
Interestingly, the two crossover temperatures of the SPM, as well as
the structure of the droplet states in the intermediate regime, bear
some resemblance with the situation in mean-field systems like the
$p$-spin model (see e.g. \cite{Castellani-Cavagna} and references
therein).

In the triangular plaquette model (TPM) we find only a single
crossover in the system at $L\sim c^{-1/\log_2 3}$. This is more in
line with the situation anticipated in BB. The same lengthscale is
again present in static multi-spin correlation functions, and in
four-point dynamical correlators.

We draw two conclusions from these results. Firstly, we are able to
verify that the procedure of BB is a suitable way to identify a mosaic
lengthscale.  In the general case it may be difficult to test the
criterion (\ref{equ:mos_len}) directly, although it is possible in the
SPM.  In less favourable situations, it is still possible to use
dynamical studies to extract the mosaic lengthscale, as we showed for
the TPM.

Secondly, we have shown that models without a finite temperature phase
transition can still be described from the mosaic perspective of KTW
and BB.  This indicates that the concept of a glassy mosaic state is
more general than the specific picture of KTW.  While the low
temperature states of the plaquette models are constructed by tiling
the plane with different metastable configurations of the spin system,
there are no well-defined droplets or domain walls in these states.
Rather, there are point-like defects that allow the system to 
interpolate smoothly between different metastable configurations.  
The crucial
difference with the KTW approach is that, while dynamics does proceed
by non-perturbative events that require co-operation over a
lengthscale that diverges at low temperatures, these events are not
related to nucleation, but to more efficient rearrangements of the
point-like defects.

In summary, typical states in the plaquette models are glassy mosaics
whose dynamics should be understood in terms of point defects with
co-operative dynamical rules. These defects are not locally conserved,
and are not related to any concept of free volume. Rather, they
represent regions in which the free energy barriers to motion are
smaller than average. In this approach, the ``activated processes'' that
dominate glassy $\alpha$-relaxation involve co-operative rearrangements
of defects over a lengthscale $\xi$. The time scales for these processes
increase with $\xi$, with a general relation 
similar to~(\ref{equ:tau_barriers}).

The purpose of this work has been to investigate rather general
arguments of KTW and BB explicitly in specific, and rather simple,
finite dimensional models.  We find good evidence for a generalised
caging effect, but our view of static and dynamic properties of the
glassy mosaic is qualitatively different from that of KTW.  The
possible links between spatial variation in the mosaic length and
dynamical heterogeneity remain as an area for further study.

\acknowledgments

We thank G. Biroli and D. Sherrington for discussions.  This work was
supported by EPSRC grants no.\ GR/R83712/01 and GR/S54074/01, and
University of Nottingham grant no.\ FEF 3024.

\begin{appendix}
\section{High temperature series in the SPM}
\label{app:pq_series}

We follow \cite{Espriu-Prats} in making a high temperature expansion
for the square plaquette model.  The partition function for a droplet
of size $L$ with fixed boundary spins is
\begin{equation}
Z_\alpha = e^{-(L-1)^2 \beta/2} \sum_{\hbox{bulk }\sigma_{xy}}
\prod_{1\leq xy \leq L-1} e^{-\beta p_{xy}/2} ,
\end{equation}
with
\begin{equation} 
p_{xy} = \sigma_{xy} \sigma_{x+1,y} \sigma_{x,y+1} \sigma_{x+1,y+1} ,
\end{equation} 
as above.  Since the $p_{xy}$ are Ising variables then this can be
written
\begin{equation}
Z_\alpha = \left(\frac{1+c}{2}\right)^{(L-1)^2}
\sum_{\hbox{bulk }\sigma_{xy}}
\prod_{1\leq xy \leq L-1} (1+zp_{xy}) ,
\end{equation}
where $z=\tanh(\beta/2)=(1-c)/(1+c)$.

Expanding the product results in a power series in $z$: there are
$2^{(L-1)^2}$ terms corresponding to states of a dual system where the
spins are the $p_{xy}$. Each state is equal to a power of $z$
multiplied by a combination of the $\sigma_{xy}$; the term vanishes on
summation over the $\sigma_{xy}$ if it contains any spin to an odd
power. Otherwise, the coefficient of the power of $z$ depends only on
boundary spins: in that case, the sum over bulk spins just contributes
a trivial factor $2^{(L-2)^2}$.

Terms containing bulk spins to even powers are formed by flipping
whole rows and columns of the dual $p_{xy}$ variables. The result is
that
\begin{eqnarray}
Z_\alpha & = & 2^{-2L+3} (1+c)^{(L-1)^2} (1/2) \times \nonumber \\ & &
\sum_{h_y,v_x\in\{0,1\}} z^{(H+V)(L-1)-2HV} \times \nonumber \\ & &
\prod_{1\leq y\leq L-1} (\sigma_{1y} \sigma_{1,y+1} \sigma_{Ly}
\sigma_{L,y+1})^{h_y} \times \nonumber \\ & & \prod_{1\leq x\leq L-1}
(\sigma_{x1} \sigma_{x+1,1} \sigma_{xL} \sigma_{x+1,L})^{v_x} ,
\end{eqnarray}
where $H=\sum_y h_y$ and $V=\sum_x v_x$; $h_y=1$ means the $p_{xy}$ in
row $y$ have been flipped. We note that the boundary spins enter only
through combinations such as $\sigma_{1y} \sigma_{1,y+1} \sigma_{Ly}
\sigma_{L,y+1}$, which gives the parity of the number of defects in
the $y$th row of the dual lattice. Further, the terms in the sum
depend on only four numbers: $H$; $V$; the number of rows for which
$h_y=1$ and $\sigma_{1y} \sigma_{1,y+1} \sigma_{Ly}
\sigma_{L,y+1}=-1$; and the number of columns for which $v_x=1$ and
$\sigma_{x1} \sigma_{x+1,1} \sigma_{xL} \sigma_{x+1,L}=-1$. Let $r$ be
the number of rows for which $\sigma_{1y} \sigma_{1,y+1} \sigma_{Ly}
\sigma_{L,y+1}=-1$ and $r'$ be the equivalent number of columns. Then
we can write
\begin{eqnarray} 
Z_\alpha &=& 2^{-2L+2} (1+c)^{(L-1)^2} 
\nonumber \\ 
&& \times 
\sum_{h=0}^r \sum_{v=0}^{r'}
\sum_{h'=0}^{L-1-r} \sum_{v'=0}^{L-1-r'} M^{(1)}_{h,v,h',v'} ,
\end{eqnarray}
with
\begin{eqnarray}
M^{(1)}_{h,v,h',v'} & = & \binc{r}{h} \binc{r'}{v} \binc{L-1-r}{h'}
 \binc{L-1-r'}{v'} \times \nonumber \\ & & (-1)^{h+v}
 z^{(H+V)(L-1)-2HV} ,
\end{eqnarray}
where now $H=h+h'$ and $V=v+v'$ since $h$ and $h'$ are the number of
rows with $h_i=1$ and odd and even defect parities respectively, and
$v$ and $v'$ are similar numbers of columns.  We denote the binomial
coefficients by $B_{n,r}\equiv n! / r!(n-r)!$

We may now sum over $v$ and $v'$, yielding
\begin{eqnarray} 
Z_\alpha &=& 2^{-2L+2} (1+c)^{(L-1)^2} \nonumber \\ && \times
\sum_{h=0}^r \sum_{h'=0}^{L-1-r} \binc{r}{h} \binc{L-1-r}{h'}
M^{(2)}_{h,v} ,
\end{eqnarray}
with
\begin{eqnarray} 
M^{(2)}_{h,h'} &=& (-1)^{h} (z^H+z^{L-1-H})^{L-r'-1} \nonumber \\ &&
\times (z^H-z^{L-1-H})^{r'} .
\end{eqnarray}

A similar procedure leads to the expression for the expectation value
of any spin, given a boundary condition. We state only the result
which depends on the coordinates of the spin $2\leq xy\leq L-1$, as
well as the total numbers of rows and columns with odd defect parity :
$r$ and $r'$. It also depends on $s$, the number of such rows with $y$
co-ordinate less than $j$ and $s'$: the number of such columns with
$x$ co-ordinate less than $i$.  (Clearly $s\leq r$ and $s'\leq r'$).

The result is that
\begin{equation} 
\langle \sigma_{xy} \rangle = \sigma_{x1} \sigma_{11} \sigma_{1y}
Z_\alpha^{-1} M_\alpha ,
\end{equation}
where
\begin{equation}
M_\alpha = \sum_{h=0}^s \sum_{h'=0}^{y-1-s} \sum_{h''=0}^{r-s} 
\sum_{h'''=0}^{L-y+s-r} M^{(3)}_{h,h',h'',h'''} ,
\end{equation}
with
\begin{eqnarray}
\lefteqn{ M^{(3)}_{h,h',h'',h'''} =} \nonumber \\ && (-1)^{h+h''}
(z^H+z^{L-1-H})^{L-x+s'-r'} \times \nonumber \\ & &
(z^H-z^{L-1-H})^{r'-s'} \times \nonumber \\ & &
(z^{H-2(h+h')}+z^{L-1-H+2(h+h')})^{x-1-s'} \times \nonumber \\ & &
(z^{H-2(h+h')}-z^{L-1-H+2(h+h')})^{s'} \times \nonumber \\ & &
B_{s',h} B_{y-1-s',h'} B_{r'-s',h''} B_{L-y+s'-r',h'''}
\end{eqnarray}

These results are still a little cumbersome, but they are exact, and
are can be evaluated computationally. The number of terms in the sum
may be as large as $\mathcal{O}(L^4)$, but this is polynomial rather
than exponential in the system size.

\section{TPM symmetries and constraints}
\label{app:triag_constraints}

Here we make some comments about the thermodynamics of finite
($L\times L$) droplets in the triangular plaquette model.  We first
identify the $2^{2L-1}$ transformations that leave the droplet energy
unchanged. Recall that
\begin{equation}
\sigma_{x+1,y} = q_{xy} \sigma_{xy} \sigma_{x,y+1} .
\end{equation}
It follows that if we specify the boundary spins $\sigma_{1y}$ and
$\sigma_{xL}$ (where $x,y=1\dots L$) then the rest of the spins are
determined by the set $\{q_{xy}\}$ (with $x,y=1\dots L-1$). Thus, the
required degenerate states for a given choice of the $\{q_{xy}\}$ can
be generated by cycling through the possible states of these two
boundaries.

To identify the constraints on the plaquette variables arising from
the boundary conditions, observe that for a ground state ($q_{xy}=1
\,\forall\, xy$) with specified spins along two boundaries
($\{\sigma_{1y},\sigma_{xL}\}$) then the state of the remaining
boundary spins are
\begin{eqnarray}
\sigma_{x1}^0 & = & \prod_{y=1}^x (\sigma_{1y})^{B_{x-1,y-1}} , \\
\sigma_{Ly}^0 & = & \prod_{y'=y}^{L-1} (\sigma_{1y'})^{B_{L-1,y'}}
\prod_{x=1}^y (\sigma_{xL})^{B_{L-1-x,L-1-y}} ,
\end{eqnarray}
where $\sigma^{B_{nr}}$ is equal to unity unless $\sigma=-1$ and
$B_{n,r}$ is odd, in which case it takes the value $-1$.

For a general configuration of the $q$'s and the boundary spins
$\sigma_{1y},\sigma_{xL}$, we have
\begin{eqnarray}
\sigma_{x1} & = & \sigma_{x1}^0 \prod_{y=1}^{x-1} \prod_{x'=1}^{x-y}
(q_{x'y})^{B_{x-x'-1,y-1}} , \\ \sigma_{Ly} & = & \sigma_{Ly}^0
\prod_{y'=0}^{L-y-1} \prod_{x=1}^{L-y'-1} (q_{y,y+y'})^{B_{L-x-1,y'}}
.
\label{equ:app_triag_constraint}
\end{eqnarray}

Multiplying both sides of each constraint by the appropriate
$\sigma^{0}$ leaves an equality between a product of boundary spins
and a product of plaquette variable. These are the $2L-3$ constraints
that determine which configurations of the excited plaquettes are
present in the partition sum.

While these expressions appear complicated, for a given droplet
boundary condition they are simply constraints of the form
\begin{equation}
q_{x_1,y_1} q_{x_2,y_2} \dots q_{x_n,y_n} = \pm 1 ,
\end{equation}
where the constrained plaquettes form parts of Sierpinski triangles.
For example, for $L=4$, the constraint coming from the boundary spin
$\sigma_{42}$ is
\begin{equation}
q_{12} q_{22} q_{32} q_{23} =
\sigma_{12} \sigma_{13} \sigma_{24} \sigma_{42} .
\end{equation}
In the SPM the constraints were of the form
\begin{equation}
\prod_{y=1}^{L-1} p_{xy} = \pm 1 .
\label{equ:app_pq_constraint}
\end{equation}
The simplicity of (\ref{equ:app_pq_constraint}) in the SPM, as
compared to (\ref{equ:app_triag_constraint}), is the reason for the
relative intractability of the partition function of the TPM.

\end{appendix}


\begin{thebibliography}{99}
\bibitem{xi-variable} We will limit our discussions to typical
 droplets.  One can also study variability in $\xi$, which is argued
 could give information about dynamic heterogeneity in the system
 \cite{Bouchaud-Biroli}.
\end{thebibliography}
\end{document}